# Finding the magnetic center of sextupole using vibrating wire technique

WU Lei(吴蕾)[1]   WANG Xiao-long(王小龙)[1,2]   LI Chun-hua(李春华)[1]   QU Hua-min(屈化民)[1,2]

[1] Institute of High Energy Physics, Chinese Academy of Sciences, Beijing 100049, China

[2] Dongguan Institute of Neutron Sciences, Dongguan 523808, China

**Abstract:**   In order to obtain stable and high-quality synchrotron radiation photon, the magnets in the storage ring of High Energy Photon Source(HEPS) need to have a stable support and precise positioning. The vibrating wire technique can be used to pre-align the quadrupoles and sextupoles on one girder with high precision to meet the extremely low emittance requirement of HEPS. This thesis introduces the measurement of magnetic center of sextupole using vibrating wire. According to the measurement results, the magnetic field distribution is consistent with theoretical expectation. And vibrating wire has achieved the purpose of measuring the magnetic center and has reached a certain precision.

## 1 Introduction

High Energy Photon Source(HEPS) is a synchrotron facility proposed by Institute of High Energy Physics which has been listed in 13th Five-Year Plan in China. It will be a 6GeV,1296m circumference third generation synchrotron radiation facility with extremely low emittance (better than 0.01mrad).Vibrating wire technique is aimed to align quadrupoles and sextupoles installed on one girder. In the preliminary R&D period, a vibrating wire measurement system was set up to verify the precision of measuring the magnetic center and the alignment effect of this technique. The detailed background was shown in ref.[1].

The measurement and alignment basic theory of vibrating wire technique has described in the ref.[2-5]. By measuring the distribution of magnetic induction intensity, vibrating wire can get the magnetic center position. The purpose of the measurement system is to measure the magnetic center of multipoles with a certain precision. The vibrating wire measurement system in HEPS is shown in Fig.1. There are one sextupole and one quadrupole installed on a multipole girder. Vibrating wire is stretched through the mechanical center of the magnet and supported by the fixed end bench and free end bench. The detailed structure of the bench has been shown in Ref.[6]. Because there are something wrong with the quadrupole, only the sextupole can be tested at present. The main measurement procedure is same both quadrupole and sextupole. And for more simple distribution of magnetic field, quadrupole measurement is easier to get a much higher precision than sextupole . This thesis mainly introduces the measurement of magnetic center of sextupole.

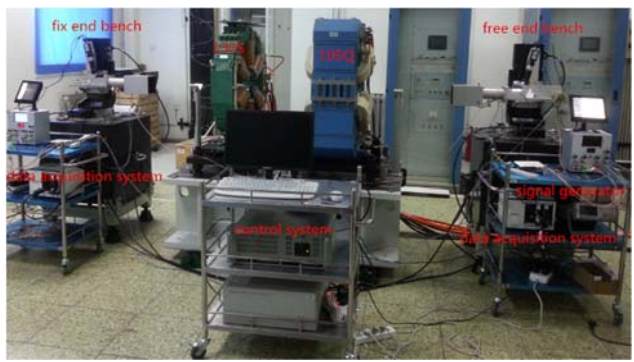

Fig. 1.   Vibrating wire system in HEPS.

## 2 Sextupole magnetic center measurement

### 2.1 The measurement procedure

The procedure of sextupole magnetic center measurement are as follows: 1) Magnet standardization: to make sure the consistency of the working state. 2) Parameters setting: parameters mainly consist of measurement direction and measurement position of the wire, amplitude and frequency of AC in the wire . By explore, the position range is set 4 mm both horizontal and vertical measurement. The frequency of AC is near to the 4[th] vibrating mode of the wire. 3) Start the automatic measurement procedure and do the data acquisition. The data acquisition of this vibrating wire measurement system mainly include two parts: one is vibrating signal of the wire and another one is the current in the wire. And the two signals should be synchronously acquired. The synchronized trigger signals are from the signal generator' SYN. 4) Analyze the acquired data and get the magnetic induction intensity distribution of the total magnetic field and the background magnetic field. The background magnetic field measurement is when the magnet without current. The total magnetic field is when the magnet with current . The total magnetic field not only includes the magnetic field generate by the measured sextupole but also the background field. The magnetic center can be found by analyzing the extreme point of parabolic distribution of magnetic induction intensity.

In a horizontal or vertical magnetic center measurement, measurement data can not only be obtained from X1 sensor installed on the fixed end bench, but also from the X2 sensor installed on the free end bench. The measurement data from this two sensors are independent and can be contrasted with each other.

**2.2 Finding sextupole magnetic center**

Fig.2 shows one horizontal magnetic center measurement result. The horizontal magnetic center is affected a lot by background field. Because the background filed distribution is linear and has a great slope in the horizontal direction for there are a quadrupole intalled on the same girder. After eliminate the background field, the vibrating wire only affected by the magnetic field of sextupole. The horizontal magnetic center position is located at -0.034mm in the vibrating wire coordinate system. The difference reaches to about 280μm with or without background correction in horizontal direction. So it is essential to do background correction. Fig.3 shows one vertical magnetic center measurement result. when measuring the vertical magnetic center, the sag correction which is about 180μm should be considered [7-9]. The background field distribution is linear but has a small slope,so the background only affected the measured vertical center by about 10μm. Even though the vertical magnetic center measurement should also do background correction. After background and sag correction, the vertical magnetic center position is located at -0.058mm.

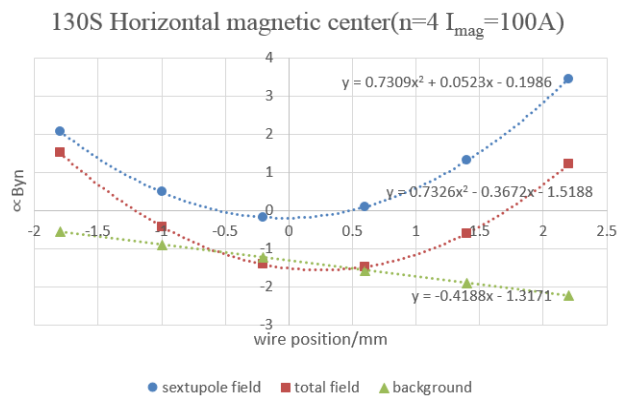

Fig. 2. Horizontal magnetic center measurement result.

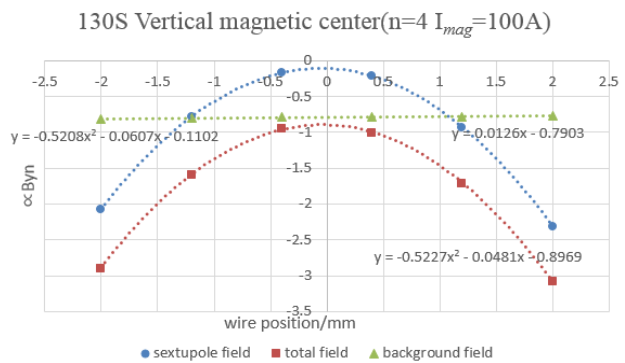

Fig. 3. Vertical magnetic center measurement result.

**2.3 The repeatability of magnetic center measurement**

In order to verify the magnetic center measurement precision, repeatability measurement is essential. 10 consecutive horizontal and vertical magnetic center measurement are shown in Fig.4 and Fig.5. From the experimental results, it shows that the consistency of horizontal magnetic center measured by X1 sensor and X2 sensor is better than 3μm, and the standard deviation of this 10 measurement is about 3μm. The consistency of vertical magnetic center measured by X1 sensor and X2 sensor is better than 6μm, the standard deviation of this 10 measurement is about 4μm. The consistency and the repeatability of vertical measurement is worse than the horizontal measurement. This may be due to the vertical measurement is more easily affected by the environment, such as the change of material temperture of magnet and girder, ground vibration and so on.

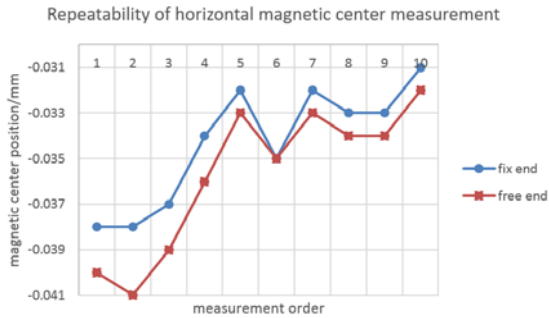

Fig. 4. 10 consecutive horizontal measurement.

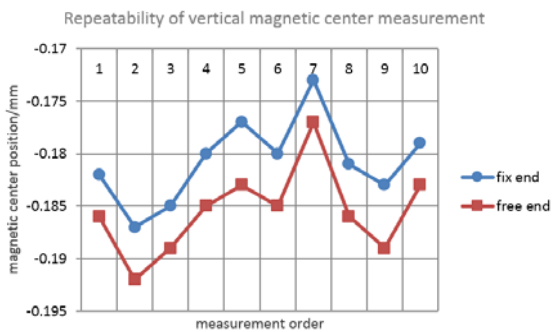

Fig. 5. 10 consecutive vertical measurement.

To further verify the precision of the magnetic center measurement, horizontal and vertical magnetic center was been tested several days. The magnetic center position is shown in Fig.6 and Fig.7. From the experiment result, the consistency of horizontal magnetic center by two sensors is better than 2μm, and the standard deviation of this 7 days measurement is about 3μm. The consistency of vertical magnetic center measured by two sensors is better than 6μm, the standard deviation of 5days measurement is about 4μm. This result is almost same with 10 consecutive measurement.

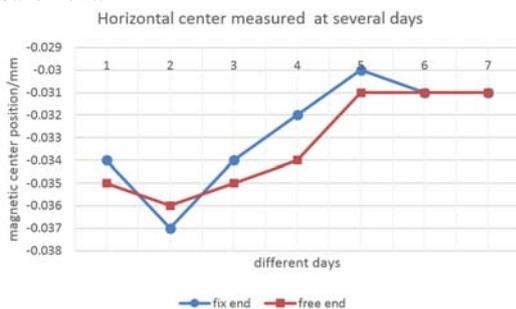

Fig. 6. Horizontal center measured at several days.

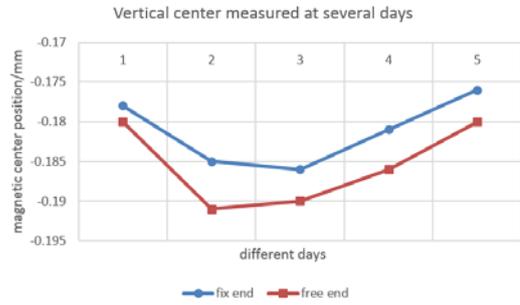

Fig. 7. Vertical center measured at several days.

### 2.4 Magnetic center at difference magnet current

Magnetic center was been measured when the magnet current is different. Fig.8 and Fig.9 is the horizontal and vertical magnetic center when the magnet current is at 80A, 100A, 120A, 140A and 160A. Considering the magnetic center measurement repeatability, the position of horizontal and vertical magnetic center measured by vibrating wire changed a little. So the magnetic center position has no relationship with the magnet current.

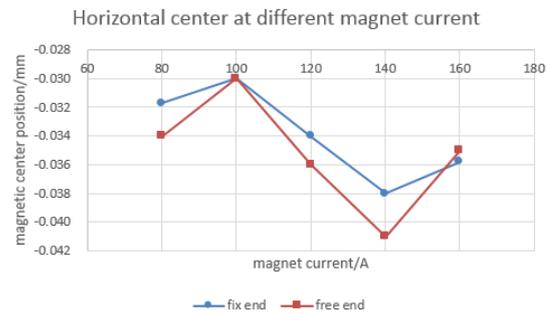

Fig. 8. Horizontal center at different magnet current.

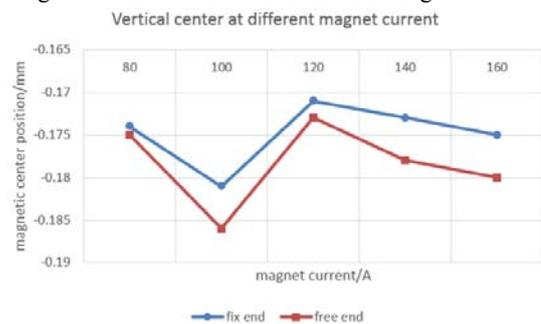

Fig. 9. Vertical center different magnet current.

### 2.5 Magnet adjustment experiment

In order to make sure the correction of this vibrating wire measurement system and the effectiveness and feasibility of alignment using vibrating wire, the magnet adjustment experiment has been done. Firstly, use vibrating wire to measure the current magnetic center of this sextupole. Secondly, adjust the sextupole in horizontal or vertical direction according to position offset of the current magnetic center. The adjustment amount is known at the monitor of the magnet position displacement sensor. Thirdly, use vibrating wire to measure magnetic center position again to see whether the magnetic center has close to the origin of the coordinate system of vibrating wire. Table 1 and table 2 show 3 times of alignment adjustment experiment results. In this experiment, the magnetic center position offsets are all smaller than the aim of 10μm after adjustment。

Table 1.  Magnet adjustment result in horizontal direction.

| Experiment sequence | Magnetic center position before alignment | Adjustment amount | Magnetic center position after alignment |
|---|---|---|---|
| 1 | 0.172mm | -0.172mm | 0.002mm |
| 2 | -0.162mm | 0.158mm | -0.001mm |
| 3 | -0.126mm | -0.126mm | -0.001mm |

Table 2.  Magnet adjustment result invertical direction.

| Experiment sequence | Magnetic center position before alignment | Adjustment amount | Magnetic center position after alignment |
|---|---|---|---|
| 1 | 0.130mm | -0.130mm | -0.003mm |
| 2 | -0.187mm | 0.174mm | -0.008mm |
| 3 | -0.020mm | 0.025mm | 0.007mm |

## 3 Conclusions

The preliminary vibrating wire system design and the magnetic center measurement experiments have gained some achievements. It has been proved the vibrating wire system designed reasonable and using vibrating wire to align the magnets installed on a multipole girder is feasible and can reach a high precision. The repeatability reaches±3μm in horizontal magnetic center measurement and ±5μm in the vertical measurement. The magnetic field distribution of quadrupole is more simple, so it will be more easily to get a higher precision than the measurement of sextupole. It is essential to measure the quadrupole when the experimental condition is satisfied in the following.